\documentclass[twocolumn,pra,showpacs]{revtex4}

\usepackage{graphicx}% Include figure files
\usepackage{dcolumn}% Align table columns on decimal point
\usepackage{bm}% bold math
\usepackage{amsmath}
\usepackage{nicefrac}

\newcommand{\ThTTN}{$^{229}$Th}
\newcommand{\ThTTNm}{$^{229m}$Th}
\newcommand{\ThIV}{Th~IV}
\newcommand{\ThIII}{Th~III}
\newcommand{\ThII}{Th~II}
\newcommand{\ThI}{Th~I}

\newcommand{\dVC}{\ensuremath{\Delta V_C}}
\newcommand{\Rms}{\ensuremath{\langle r^2 \rangle}}
\newcommand{\Rrms}{\ensuremath{r_{rms}}}
\newcommand{\dRms}{\ensuremath{\Delta \langle r^2 \rangle}}
\newcommand{\dRrms}{\ensuremath{\Delta r_{rms}}}
\newcommand{\dQ}{\ensuremath{\Delta Q_0}}

\newcolumntype{d}{D{.}{.}{-1}}    %Align on decimal point
\newcolumntype{b}{D{(}{\ (}{-1}}  %Align on opening bracket of errors; put space 

\newcommand{\cm}{\text{cm$^{-1}$}}

\newcommand{\GHzfm}{GHz/fm$^2$}
\newcommand{\E}[1]{\ensuremath{\times 10^{#1}}}
\newcommand{\rtw}{\ensuremath{\rightarrow}}

\newcommand{\Cite}[1]{\mbox{Ref.~\cite{#1}}}
\newcommand{\Cites}[1]{\mbox{Refs.~\cite{#1}}}
\newcommand{\eref}[1]{(\ref{#1})}
\newcommand{\Eref}[1]{Eq.~(\ref{#1})}
\newcommand{\Tref}[1]{Table~\ref{#1}}

\newcommand{\Sec}[1]{Section~\ref{#1}}

\begin{document}

\title{A proposed experimental method to determine $\alpha$-sensitivity of splitting between ground and 7.6 eV isomeric states in $^{229}$Th}
%\title{Determining changes in the nuclear charge radius of Th isomers}

\author{J. C. Berengut}
\affiliation{School of Physics, University of New South Wales, Sydney 2052, Australia}
\author{V. A. Dzuba}
\affiliation{School of Physics, University of New South Wales, Sydney 2052, Australia}
\author{V. V. Flambaum}
\affiliation{School of Physics, University of New South Wales, Sydney 2052, Australia}
\author{S. G. Porsev}
\affiliation{School of Physics, University of New South Wales, Sydney 2052, Australia}
\affiliation{Petersburg Nuclear Physics Institute, Gatchina, 188300, Russia}

\date{16 April 2009}

\begin{abstract}

The 7.6 eV electromagnetic transition between the nearly degenerate ground state and first excited state in the \ThTTN\ nucleus may be very sensitive to potential changes in the fine-structure constant, $\alpha = e^2/\hbar c$. However, the sensitivity is not known, and nuclear calculations are currently unable to determine it. We propose measurements of the differences of atomic transition frequencies between thorium atoms (or ions) with the nucleus in the ground state and in the first excited (isomeric) state. This will enable extraction of the change in nuclear charge radius and electric quadrupole moment between the isomers, and hence the $\alpha$-dependence of the isomeric transition frequency with reasonable accuracy.

\end{abstract}


\maketitle

\section{\label{sec:intro} Introduction}

The isotope \ThTTN\ has the lowest known excited state in nuclei; recent measurements show that the $3/2^+$ state lies just 7.6 eV above the $5/2^+$ ground state~\cite{beck07prl}. The width of this level is estimated to be about $10^{-4}$~Hz~\cite{tkalya00prc} which may explain why it is so hard to find the direct radiation in this very weak transition. Nevertheless, the frequency is within the range of lasers, and it has been proposed to use this narrow nuclear transition as possible reference for an optical clock of very high accuracy~\cite{peik03epl}. Additionally, this transition could be a sensitive probe of possible variation of fundamental constants~\cite{flambaum06prl} because the near degeneracy of these isomers is a result of cancelation between very large energy contributions (order of MeV). Since these contributions would have different dependences on fundamental constants, any variation would be enhanced in the transition frequency. In \Cite{flambaum06prl}, the relative effects of variation of $\alpha$ and the dimensionless strong interaction parameter, $m_q/\Lambda_{QCD}$ were estimated to be enhanced by 5 orders of magnitude.

An enhancement to $\alpha$-sensitivity of this magnitude would have very important consequences for laboratory searches of $\alpha$-variation. Because the isomeric \ThTTN\ resonance has a narrow linewidth and an extraordinary insensitivity to external perturbations, an optical clock utilising this reference may have very high accuracy and high immunity from systematic frequency shifts~\cite{peik03epl}. By comparing this ``nuclear clock'' frequency with that of any other narrow optical or microwave transition (e.g. the Cs or Hg$^+$ frequency standards) one can test variation of fundamental constants. Coupled with the enhancement in sensitivity, such a set up would be the most sensitive laboratory probe of $\alpha$-variation to date, possibly gaining several orders-of-magnitude improvement over the current limits of
\mbox{$\dot\alpha/\alpha = (-1.6 \pm 2.3)\times 10^{-17}$~year$^{-1}$~\cite{rosenband08sci}}.

The sensitivity of the transition frequency to variation of $\alpha$ can be expressed as
\begin{equation}
	\delta \omega = \dVC\ \frac{\delta\alpha}{\alpha}, \quad
	\frac{\delta\omega}{\omega} = K \frac{\delta\alpha}{\alpha}
\end{equation}
where \dVC\ is the difference in Coulomb energies between the two isomers, and $K$ is the enhancement factor: \mbox{$K = \dVC/\omega$}. Since the Coulomb energy of this nucleus is of order $10^9$~eV, even a relatively small variation in $V_C$ could produce a large enhancement. For $\dVC = 100$~keV and $\delta\alpha/\alpha = 10^{-16}$, $\delta\omega = 10^{-11}~\textrm{eV} = 2.4\E{3}$~Hz which is 4 orders of magnitude larger than the limits placed on shifts in atomic transitions in \Cite{rosenband08sci}.

However, different nuclear calculations give wildly different values for \dVC. \Cites{hayes07plb,hayes08prc} claim that both isomers have identical deformations and therefore the same Coulomb energies to within roughly 30~keV (corresponding to $K \lesssim 4000$). \Cite{he08jpg} gives a value of 30~keV, while the calculations of \cite{litvinova09arxiv} give values in the range \mbox{-300 keV $< \dVC\ <$ 450 keV}, depending on particulars of the model used. Lastly, \Cite{flambaum09epl} uses Nilsson wave functions to show that the value of \dVC\ as a function of deformation changes from 1.5 MeV at zero deformation down to -0.5 MeV at $\delta = 0.3$. They conclude that a very small value of the Coulomb energy shift seems improbable.

In this paper we propose a different method for extracting sensitivity to $\alpha$-variation using direct laboratory measurements of the change in nuclear mean-square charge radius, \dRms, and electric quadrupole moment, \dQ, between the isomer and the ground state nucleus. In \Sec{sec:geometry} we present a simple geometric model of the nucleus to relate the observable nuclear parameters to \dVC\ and hence $K$. We show that this model is self-consistent by comparing to the nuclear calculations of \Cite{litvinova09arxiv}.

Once the change in Coulomb energy has been measured, the change in nuclear energy $\Delta E_N$ will be known also, since they almost cancel for this transition: $\Delta E_N + \dVC = 7.6$~eV. This change in nuclear energy can be interpreted in terms of variation of the dimensionless ratio $m_q/\Lambda_{QCD}$ where $m_q$ is the light quark mass and $\Lambda_{QCD}$ is the pole in the running strong coupling constant. Variation of this ratio will also be enhanced in the \ThTTN\ transition \cite{flambaum06prl}; calculations may be found in \cite{flambaum09prc,he08jpg}.

The change in mean-square nuclear radius can be extracted using the isomeric field shift for an atomic transition. In principle, any transition in any \ThTTN\ ion or the neutral atom can be used. There are two approaches. The first is entirely empirical: by combining the measurements of isomeric shifts and isotopic shifts for the same transition, one can extract the ratio of \dRms\ for the isomer to the isotopic change in mean-square radius. The second approach does not require the additional measurement of isotope shift, but it does require high-precision atomic calculations. We provide more details and necessary calculations in \Sec{sec:field}.

To extract the change in nuclear quadrupole moment, the hyperfine structure may be used. The hyperfine structure constant $B$, which can be determined experimentally, is proportional to the quadrupole moment $Q$. Therefore one must measure $B$ for both the ground state and isomeric \ThTTN. The value of $Q$ for the ground state is known to within 20\%; better accuracy can be obtained using the calculations presented in \Sec{sec:quadrupole}.

%In \Sec{sec:field} we present calculations of field-shift constants for \ThIV\ since it is alkali-like and is amenable to laser cooling and trapping\cite{peik03epl}. Note, however, that if one obtains an isotope shift for the same transition there is no need for high precision atomic calculations. Field-shift constants for \ThII\ and \ThIII\ are also presented. In \Sec{sec:quadrupole} the calculations required to extract the change in nuclear quadrupole moment from the hyperfine splitting of \ThIV\ are presented. Again, this is unnecessary if it can be reliably measured by direct means.

The radiative lifetime of the metastable \ThTTNm\ nucleus is estimated to be a few hours~\cite{tkalya00prc}, however this may be reduced if the energy of the excited state exceeds the ionization potential since an electron autoionization channel may open. The successive ionization energies of thorium ions are~\cite{crchandbook}: 6.3~eV~(\ThI), 11.9~eV~(\ThII), 20.0~eV~(\ThIII), 28.8~eV~(\ThIV). Therefore, the atomic experiments are likely to be easier for ionized thorium since the ionization energies exceed the excitation energy. In fact, \ThIV\ may be the best choice since it is alkali-like and is amenable to laser cooling and trapping~\cite{peik03epl}. This ion has the additional advantage that calculations are likely to be more accurate, although we stress that we can obtain reasonable accuracy with any ion that experimentalists may find convenient.

\section{\label{sec:geometry} Geometrical Nuclear Model}

In this section we use a simple geometric model to relate the Coulomb energy of a nucleus to the experimentally observable mean-square charge radius and electric quadrupole moment. We assume that both the ground-state nucleus and the lowest-energy isomer are uniform, hard-edged, prolate ellipsoids. Let $a$ and $c$ be the semi-minor and semi-major axes, respectively. We define $R_0$ as the equal-volume spherical radius and $\zeta$ as the eccentricity, so that
\begin{equation}
	R_0^3 = a^2 c \quad\textrm{and}\quad
	\zeta^2 = 1 - \frac{a^2}{c^2}\ .
\end{equation}

We can extract from experiment the mean-square radius and intrinsic electric quadrupole moment (see Sections~\ref{sec:field} and \ref{sec:quadrupole}) defined as
\begin{eqnarray}
\Rms &=& \int r^2 \rho (r) d^3 r \\
 Q_0 &=& \int r^2 \left( 3\cos^2(\theta) - 1 \right) \rho (r) d^3 r
\end{eqnarray}
where $\rho(r)$ is the electric charge density normalised to unity.
The intrinsic quadrupole moment is related to the laboratory quadrupole moment of the ground rotational mode by (see, e.g. \cite{segre77book})
\begin{equation}
\label{eq:Qlab}
Q_{\textrm{lab}} = Z Q_0 \frac{I(2I-1)}{(I+1)(2I+3)}\ .
\end{equation}
For our hard-shell prolate nucleus, one finds
\begin{equation*}
\Rms = \frac{1}{5}(2a^2 + c^2) \quad\textrm{and}\quad
 Q_0 = \frac{2}{5}(c^2 - a^2)\ .
\end{equation*}

We wish to express the Coulomb energy in terms of these measurable quantities. Using formulas presented in \cite{hasse88book} we find
\begin{eqnarray}
V_C &=& \frac{3}{5}\frac{(Ze)^2}{R_0}
       \frac{(1+\zeta^2)^{1/3}}{2\zeta} \log \frac{1+\zeta}{1-\zeta} \nonumber \\
 &\approx& \frac{3}{5}\frac{(Ze)^2}{R_0}
        \left( 1 - \frac{1}{45}\zeta^4 + O(\zeta^6) \right)
\end{eqnarray}
and in terms of \Rms and $Q_0$ we finally obtain
\begin{equation}
\label{eq:Vc}
V_C = \left(\frac{3}{5}\right)^{\nicefrac{3}{2}} \frac{(Ze)^2}{\Rms^{1/2}}
     \left( 1 + \frac{3}{40}\frac{Q_0^2}{\Rms^2} - \frac{1}{56}\frac{Q_0^3}{\Rms^3}
            + ... \right)
\end{equation}
With this equation we can extract \dVC\ if we know \dRms\ and \dQ\ between the \ThTTN\ isomers. Note that $V_C$ and \dVC\ are vastly more sensitive to changes in \Rms\ than $Q_0$.

To estimate the effect of skin thickness, we use a spherical Fermi distribution model:
\begin{equation}
\rho(r) = \rho_0 \left( 1 + \exp \frac{r-C}{z} \right)^{-1}\ .
\end{equation}
The Coulomb energy is
\begin{equation}
V_C = \left(\frac{3}{5}\right)^{\nicefrac{3}{2}} \frac{(Ze)^2}{\Rms^{1/2}}
     \left( 1 + 8.379 \frac{z^3}{\Rms^{3/2}} + ... \right)
\end{equation}
and one sees that $V_C$ is not sensitive to the skin thickness parameter $z$.

With the current data for \ThTTN, $\Rrms = 5.6807 \pm 0.0509$ fm~\cite{angeli04adndt} and \mbox{$Q_{\textrm{lab}} = 4.3 \pm 0.9$ $e$b~\cite{stone05adndt}}, we obtain $Q_0 = 13.4$, an eccentricity $\zeta^2 = 0.440$,
and Coulomb energy $V_C = 967$~MeV. We estimate $z = 0.5$~fm. In this case the change in Coulomb energy can be expressed
\begin{equation}
\label{eq:dVC}
\frac{\dVC}{\textrm{(MeV)}}
   = -506\,\frac{\dRms}{\Rms} + 23\,\frac{\dQ}{Q_0} + 17\,\frac{\Delta z}{z}
\end{equation}
from which the sensitivity of the transition to $\alpha$-variation is easily deduced. Note that the contribution of $\Delta z$ is small.

As a consistency check, we have recalculated \dVC\ using the values of \Rrms, \dRrms, $Q_0$, and \dQ\ calculated in \Cite{litvinova09arxiv}; this is shown in \Tref{tab:VC_consistency}. That we are able to reproduce their results shows the validity of geometrical model. The differences seen in the SIII entries of \Tref{tab:VC_consistency} (last two columns) are probably due to lack of numerical precision. If measurements of \dRrms\ and \dQ\ are made accurately, the model should suffice even when the measurable nuclear parameters are small.

\begin{table}
\caption{The values of \Rrms, $Q_0$, \dRrms, \dQ, and $V_C$ are reproduced from \Cite{litvinova09arxiv} and used to calculate the value of \dVC\ shown in the last line using our simple geometrical model. SkM$^*$ and SIII refer to two different energy functionals, while HF and HFB refer to Hartree-Fock and Hartree-Fock-Bogoliubov, the latter includes pairing correlations; for details see \Cite{litvinova09arxiv}.}
\label{tab:VC_consistency}
\begin{tabular}{crrrr}
\hline \hline
  & \multicolumn{2}{c}{SkM$^*$} &  \multicolumn{2}{c}{SIII} \\
  & \multicolumn{1}{c}{HF} & \multicolumn{1}{c}{HFB}
  & \multicolumn{1}{c}{HF} & \multicolumn{1}{c}{HFB} \\
\hline
\Rrms\ (fm)\footnotemark[1] 
  & 5.7180 & 5.7078 & 5.7817 & 5.7769 \\
$Q_0$ (fm$^2$)\footnotemark[1]
  & 9.5461 & 9.3717 & 9.3542 & 9.1643 \\
$\Delta \Rrms$ (fm)\footnotemark[1]
  & -0.0038 & 0.0039 & 0.0000 & -0.0005 \\
\dQ\ (fm$^2$)\footnotemark[1]
  & -0.1824 & 0.2756 & -0.0339 & -0.0495 \\
$V_C$ (MeV)\footnotemark[2]
  & 924 & 925 & 912 & 912 \\
\\
\dVC\ (MeV)\footnotemark[2]
  & 0.451 & -0.307 & -0.098 & 0.001 \\
\dVC\ (MeV)
  & 0.419 & -0.327 & -0.036 & 0.029 \\
\hline\hline
\end{tabular}
\footnotetext[1]{From \Cite{litvinova09arxiv}, Table~II.}
\footnotetext[2]{From \Cite{litvinova09arxiv}, Table~I.}
\end{table}

\section{\label{sec:field} Mean-square Radius}

In the previous section we showed that \dRms\ is the most important
quantity for determining \dVC\ and hence sensitivity to
$\alpha$-variation. In this section we show how \dRms can be extracted from the isomeric shift of any atomic transition, obtained by comparing \ThTTN\ and \ThTTNm. This is similar to the usual isotope shift.

The shift in energy of any transition in an isotope with mass number
$A'$ with respect to an isotope with mass number $A$ can be expressed
as
\begin{equation}
\label{eq:is}
\Delta \nu^{A', A} = \left( k_{\rm NMS} + k_{\rm SMS} \right)  \left(
    \frac{1}{A'} - \frac{1}{A} \right) + F \dRms
    ^{A', A} \ .
\end{equation}
Here the first term is the ``mass shift'' due to the finite mass of
the nucleus and the second term is the ``volume'' or ``field'' shift
due to the finite size of the nuclear charge distribution (see,
e.g.~\cite{berengut03pra}). In the case of the isomeric shift that we
are interested in, the mass shift vanishes since isomers have equal
mass. Thus in order to extract \dRms\ from a measurement of isomeric
shift $\Delta\nu^m$ for an atomic transition we need simply divide by
the field-shift constant $F$:
\begin{equation}
\label{eq:F}
\Delta\nu^m = F\, \dRms\ .
\end{equation}
These may be calculated or extracted from known isotope shifts.

In Tables \ref{tab:ThIV}, \ref{tab:ThIII}, and \ref{tab:ThII} we present calculated field shift constants for transitions in several ions of Th. In \Tref{tab:ThIV} we have included an estimated size of the isomeric shift, $\delta\nu^m$, assuming that $\dRrms = 0.004$~fm, which is the magnitude of the largest shifts in \cite{litvinova09arxiv} (from the SkM$^*$ nuclear energy functionals).

%----------------------------------------------------------------------
\begin{table}
\caption{Calculated energies and field shift constants of transitions in \ThIV. The last column shows expected ``order of magnitude'' isomeric shifts in \ThTTN, assuming $|\dRrms| = 0.004$~fm, however the actual shift could differ by an order of magnitude. All transitions are to the $5f_{5/2}$ ground state.}
\label{tab:ThIV}
\begin{tabular}{lrrbc}
\hline \hline
  & \multicolumn{2}{c}{$\omega$~(\cm)}
  & \multicolumn{1}{c}{$F$}
  & \multicolumn{1}{c}{$|\delta\nu^m|$} \\
\multicolumn{1}{c}{Level}
  & \multicolumn{1}{c}{Expt.}
  & \multicolumn{1}{c}{Calc}
  & \multicolumn{1}{c}{(\GHzfm)}
  & \multicolumn{1}{c}{(GHz)} \\
\hline
$5f_{7/2}$ &  4325 &  4899 &   2(2) & 0.09 \\
$6d_{3/2}$ &  9193 & 11721 &  33(8) & 1.4 \\
$6d_{5/2}$ & 14486 & 17534 &  35(8) & 1.5 \\
$7s_{1/2}$ & 23131 & 24740 & 146(4) & 6.3 \\
$7p_{1/2}$ & 60239 & 63051 &  57(3) & 2.5 \\
$7p_{3/2}$ & 73056 & 76319 &  49(2) & 2.1 \\
\hline\hline
\end{tabular}
\end{table}
%----------------------------------------------------------------------

We calculate the field shift constants $F$ using methods developed in previous
works~\cite{berengut03pra}. Briefly, we perform an energy calculation
several times, modifying the nuclear radius in our codes. $F$ is
extracted from the  gradient: $F = dE/d\Rms$ at $\Rrms = 5.6807$~fm.

Calculations of the energies are slightly different for a
single-valence-electron ion (\ThIV) and for two- and three-valence-electron
ions (\ThIII\ and \ThII). In the former case we use the
correlation-potential method developed in Ref.~\cite{dzuba87jpb}. The
second-order correlation correction potential $\hat \Sigma^{(2)}$ is
used to calculate Brueckner orbitals for the states of the valence
electron. This techniques takes into account dominating relativistic and
correlation effects and leads to good agreement between theoretical
and experimental energies as illustrated in Table~\ref{tab:ThIV}.

For ions with two and three valence electrons we use the combination
of the many-body perturbation theory and the configuration interaction
technique (CI+MBPT, Refs.~\cite{dzuba96pra,dzuba98pra}). The same
single-electron correlation correction operator $\hat \Sigma_1$ is
used for all three ions, including the single-electron ion
\ThIV. However, an extra two-electron correlation correction operator
$\hat \Sigma_2$ is needed for ions with more than one valence electron (see
Refs.~\cite{dzuba96pra,dzuba98pra} for details). The accuracy of these
calculations is also high, as is illustrated in Tables~\ref{tab:ThIII} and \ref{tab:ThII}.

%----------------------------------------------------------------------

\begin{table}
\caption{Calculated energies, $\omega$~(\cm), field shift constants, $F$~(\GHzfm), and isotope shifts, $\delta\nu^{232,230}$~($10^{-3}$\cm), of some transitions in \ThIII. All transitions are to the $5f6d \ ^3$H$^o_4$ ground state. Note that, while we believe the $6d^2\ ^3$F$_3$, $6d^2\ ^3$F$_4$, and $6d7s\ ^3$D$_3$ transitions are accurate, the others are estimates only.}
\label{tab:ThIII}
\begin{tabular}{lcrrrr}
\hline \hline
 \multicolumn{2}{c}{Level}
  & \multicolumn{2}{c}{$\omega$~(\cm)}
  & \multicolumn{1}{c}{$F$}
  & \multicolumn{1}{r}{$\delta\nu^{232,230}$} \\
\multicolumn{1}{l}{Term}
  & \multicolumn{1}{c}{$J$}
  & \multicolumn{1}{c}{Exp.}
  & \multicolumn{1}{c}{Calc}
  & %\multicolumn{1}{c}{(\GHzfm)}
  & \multicolumn{1}{r}{Calc.\footnotemark[1]} \\
%  & \multicolumn{1}{c}{Calc.\footnotemark[2]} \\
\hline
$6d^2\ ^3$F & 3 &  4056 &  4023 &  24 &  165 \\%&  149 \\
$6d^2\ ^3$F & 4 &  6538 &  6795 &  22 &  147 \\%&  133 \\
$6d7s\ ^3$D & 3 &  9954 &  9204 & 118 &  804 \\%&  725 \\
$6d^2\ ^1$G & 4 & 10543 & 11051 &   8 &   56 \\%&   51 \\
$5f^2\ ^3$H & 4 & 15149 & 13358 & -11 &  -77 \\%&  -70 \\
$5f^2\ ^3$H & 5 & 17887 & 16068 & -20 & -136 \\%&  -123 \\

$5f^2\ ^3$F & 3 & 20840 & 19080 & -18 & -122 \\%& -110 \\
$5f^2\ ^3$F & 4 & 21784 & 20366 & -15 & -101 \\%&  -91 \\
$5f^2\ ^1$G & 4 & 25972 & 25269 &  10 &  -66 \\%&  -60 \\
$5f7p\ (\frac{5}{2},\frac{1}{2})$ &              
                 3 & 33562 & 33402 &  13 &  92 \\%&   83 \\
$5f7p\ (\frac{7}{2},\frac{1}{2})$ &              
                 3 & 38432 & 38617 &  15 &  101 \\%&   91 \\
\hline\hline
\end{tabular}
\footnotetext[1]{$\Delta\langle r^2 \rangle=0.205$ fm$^2$, from \Cite{angeli04adndt}}
%\footnotetext[2]{$\Delta\langle r^2 \rangle=0.185$ fm$^2$, best fit value.}
\end{table}
%----------------------------------------------------------------------

\begin{table}
\caption{Calculated energies, $\omega$~(\cm), field shift constants, $F$~(\GHzfm), and isotope shifts, $\delta\nu^{232,230}$~($10^{-3}$\cm), of some transitions in \ThII. All transitions are to the $6d^27s \ J=3/2$ ground state.}
\label{tab:ThII}
\begin{tabular}{lccrrrrrr}
\hline \hline
 \multicolumn{3}{c}{Level}
  & \multicolumn{2}{c}{$\omega$~(\cm)}
  & \multicolumn{1}{c}{$F$}
  & \multicolumn{3}{c}{$\delta\nu^{232,230}$} \\
\multicolumn{2}{c}{Configuration}
  & \multicolumn{1}{c}{$J$}
  & \multicolumn{1}{c}{Exp.}
  & \multicolumn{1}{c}{Calc}
  &% \multicolumn{1}{c}{(\GHzfm)}
  & \multicolumn{1}{c}{Exp.} 
  & \multicolumn{1}{c}{Calc.\footnotemark[1]} 
  & \multicolumn{1}{c}{Calc.\footnotemark[2]} \\
\hline
$5f7s^2$ & $^2$F$^o$ & 5/2 &  4490 &  4856 &   4 &   54 &   47 &   43 \\
$5f6d7s$ & $^4$F$^o$ & 3/2 &  6691 &  7487 & -53 & -362 & -401 & -362 \\
$5f6d7s$ & $^4$F$^o$ & 5/2 &  7331 &  8325 & -53 & -365 & -405 & -365 \\
$5f6d7s$ & $^4$G$^o$ & 5/2 &  9585 & 10045 & -55 & -375 & -406 & -366 \\
$5f6d7s$ & $^4$H$^o$ & 5/2 & 10673 & 12168 & -53 & -361 & -406 & -367 \\
$5f6d7s$ & $^2$D$^o$ & 3/2 & 11576 & 13054 & -54 & -367 & -408 & -368 \\
$5f6d7s$ & $^4$D$^o$ & 1/2 & 11725 & 12897 & -67 & -456 & -460 & -415 \\
$5f6d7s$ & $^2$F$^o$ & 5/2 & 12472 & 14564 & -58 & -399 & -463 & -418 \\
$5f6d7s$ & $^4$F$^o$ & 3/2 & 12902 & 14233 & -58 & -395 & -444 & -400 \\
$5f6d7s$ & $^4$G$^o$ & 1/2 & 14102 & 15853 & -79 & -539 & -610 & -550 \\

%$5f6d7s$ & $^4$D$^o$ & 1/2 & 11725 & 12897 & -67 & -456 & -460 & -415 \\
%$5f6d7s$ & $^4$G$^o$ & 1/2 & 14102 & 15853 & -89 & -539 & -610 & -550 \\
%$5f6d7s$ & $^4$P$^o$ & 1/2 & 15324 & 17069 & -57 &      & -393 & -354 \\
%$5f6d7s$ & $^4$H$^o$ & 1/2 & 17838 &       &     &      &      &
                                                            
%$5f6d7s$ & $^4$F$^o$ & 3/2 &  6691 &  7487 & -59 & -362 & -401 & -362 \\
%$5f6d7s$ & $^2$D$^o$ & 3/2 & 11576 & 13054 & -60 & -367 & -408 & -368 \\
%$5f6d7s$ & $^4$F$^o$ & 3/2 & 12902 & 14233 & -65 & -395 & -444 & -400 \\
%$5f6d7s$ & $^4$D$^o$ & 3/2 & 15145 & 17093 &     &      &      &
%$5f6d7s$ & $^2$D$^o$ & 3/2 & 15711 & 17597 &     &      &      &
%$5f6d^2$ & $^4$I$^o$ & 3/2 & 17122 & 19503 &     &      &      &
                                                             
%$5f7s^2$ & $^2$F$^o$ & 5/2 &  4490 &  4856 &   7 &   54 &   47 &   43 \\
%$5f6d7s$ & $^4$F$^o$ & 5/2 &  7331 &  8325 & -59 & -365 & -405 & -365 \\
%$5f6d7s$ & $^4$G$^o$ & 5/2 &  9585 & 10045 & -59 & -375 & -406 & -366 \\
%$5f6d7s$ & $^4$H$^o$ & 5/2 & 10673 & 12168 & -59 & -361 & -406 & -367 \\
%$5f6d7s$ & $^2$F$^o$ & 5/2 & 12472 & 14564 & -68 & -399 & -463 & -418 \\

\hline\hline
\end{tabular}
\footnotetext[1]{$\Delta\langle r^2 \rangle=0.205$ fm$^2$, from \Cite{angeli04adndt}}
\footnotetext[2]{$\Delta\langle r^2 \rangle=0.185$ fm$^2$, best fit value.}
\end{table}
%----------------------------------------------------------------------

For \ThII\ there are experimental isotope shifts available~\cite{actinidesbook} and we compare them with our calculations in \Tref{tab:ThII}. Note that the mass shift has been ignored here: while $k_{\rm SMS}$ is difficult to evaluate accurately, $k_\mathrm{NMS}$ is easily extracted from the transition frequency and is proportional to it. If we assume that $k_\mathrm{NMS}$ and $k_\mathrm{SMS}$ are of the same order, then $k_\mathrm{NMS} (1/232 - 1/230) \approx 2\E{-8}\,\nu$ is negligible. The second-last column of \Tref{tab:ThII} is a calculation with \mbox{$\dRms^{232,230} = 0.205\,(30)$ fm$^2$}~\cite{angeli04adndt}. The last column gives values of the isotope shift with \mbox{$\dRms^{232,230} = 0.185$}: this is the value that gives the best fit of our calculated isotope shifts to the experimental data.

The field shift constant is generally larger for transitions involving a change in the $s$-wave configuration, e.g. $5f_{5/2} \rtw 7s_{1/2}$ transition in \ThIV\ and the $5f6d \ ^3\textrm{H}^o_4 \rtw 6d7s\ ^3\textrm{D}_3$ transition in \ThIII. Measurement of the isomeric shift may be easier for these cases. However if there are good reasons to use transitions with smaller shifts (e.g. the higher-energy transitions in \ThIII), then we recommend the experimentalists contact us for more precise values of the constants. Again we stress that these constants may be extracted from measured isotope shifts with accuracy limited by knowledge of the isotopic change in mean-square radius, $\dRms^{A', A}$.

\section{\label{sec:quadrupole} Electric Quadrupole Moment}

Although we have shown in \Sec{sec:geometry} that the change in Coulomb energy of the 7.6 eV transition in the \ThTTN\ nucleus is far more sensitive to \Rms\ than $Q_0$, \dQ\ could still be important if \dRms\ is found to be very small. Fortunately \dQ\ can be extracted from measurements of the hyperfine structure of the isomers by using \eref{eq:Qlab} and noting that $Q_\textrm{lab}$ is proportional to the electric-quadrupole hyperfine-structure constants $B$.

Since the electric quadrupole moment of the ground state \ThTTN\ nucleus is known to about 20\% accuracy (\mbox{$Q_\textrm{lab}=4.3(9)~e$b~\cite{stone05adndt}}), \dQ\ can be extracted by measuring the electric quadrupole hyperfine structure of both isomers. This can be done for any states of any thorium ion or neutral atom and no atomic calculations are needed for the interpretation of the results.

If better than 20\% accuracy is required, the values of $Q_\textrm{lab}$ can be found by comparision of the calculated and measured $B$. Calculations with this level of accuracy for many-valence-electrons are difficult, but can be
performed if required. In this work we present the calculations of $B$
for the single-valence-electron ion \ThIV. The calculations are done
with the correlation potential method which takes into account dominating correlation corrections~\cite{dzuba87jpb}. The constant $B$ for
a particular valence state $v$ is found as a matrix element
\begin{equation}
  B_v = A\, \langle \psi_v^{Br} || \hat F + \delta V || \psi_v^{Br}
  \rangle,
\label{eq:Bhfs}
\end{equation}
where $A$ is a numerical constant, $\psi_v^{Br}$ is the Brueckner
orbital for the valence state $v$, $\hat F$ is the operator of
the nuclear electric quadrupole moment and $\delta V$ is the
correction to the atomic self-consistent potential due to the effect
of nuclear quadrupole electric field on atomic electrons.
The same Brueckner orbitals are used as in the previous section. The
results are presented in Table~\ref{tab:Bhfs}: accuracy is
expected to be at the level of a few per cent.  

\begin{table}
\caption{Calculated electric-quadrupole hyperfine-structure constants $B$ for some low energy states of \ThIV. In the last column, the nuclear electric quadrupole moment $Q$ is taken to be 4.3~b.}
\label{tab:Bhfs}
\begin{tabular}{lrr}
\hline \hline
\multicolumn{1}{c}{Level}
  & \multicolumn{2}{c}{$B$ ~(MHz)} \\
\hline
$5f_{5/2}$ &  $740\,Q$ & \ 3180 \\
$5f_{7/2}$ &  $860\,Q$ & \ 3700 \\
$6d_{3/2}$ &  $690\,Q$ & \ 2970 \\
$6d_{5/2}$ &  $860\,Q$ & \ 3700 \\
$7p_{3/2}$ & $1810\,Q$ & \ 7790 \\
\hline\hline
\end{tabular}
\end{table}

\section{Conclusion}

We have presented a simple geometrical model which allows one to calculate changes in the Coulomb energy between the different isomers given small changes in mean-square radius and quadrupole moment; with current data the change is given by \Eref{eq:dVC}. These parameters can be obtained by measurement of the atomic spectra of \ThTTN\ and its isomer. From the change in Coulomb energy, the sensitivity of the isomeric transition frequency to $\alpha$-variation can easily be deduced.

Two approaches have been proposed for measuring the change in mean-square charge radius: in the first the isotope shift must be measured in conjunction with the isomeric shift. In the second approach measurement of an isotope shift is not needed, but atomic calculations are required to interpret these measurements. We have shown that we can calculate the relevant parameters: namely $F$ for extracting \dRms\ (Eq.~\ref{eq:F}) and $B$ for extracting \dQ\ (Eq.~\ref{eq:Bhfs}). We recommend that experimentalists contact us for more accurate calculations for the atomic transitions that they intend to exploit.

\section*{Acknowledgments}

This work is supported by the Australian Research Council, Marsden Grant, and the NCI National Facility.

\bibliography{references}

\begin{thebibliography}{21}
\expandafter\ifx\csname natexlab\endcsname\relax\def\natexlab#1{#1}\fi
\expandafter\ifx\csname bibnamefont\endcsname\relax
  \def\bibnamefont#1{#1}\fi
\expandafter\ifx\csname bibfnamefont\endcsname\relax
  \def\bibfnamefont#1{#1}\fi
\expandafter\ifx\csname citenamefont\endcsname\relax
  \def\citenamefont#1{#1}\fi
\expandafter\ifx\csname url\endcsname\relax
  \def\url#1{\texttt{#1}}\fi
\expandafter\ifx\csname urlprefix\endcsname\relax\def\urlprefix{URL }\fi
\providecommand{\bibinfo}[2]{#2}
\providecommand{\eprint}[2][]{\url{#2}}

\bibitem[{\citenamefont{Beck et~al.}(2007)\citenamefont{Beck, Becker,
  Beiersdorfer, Brown, Moody, Wilhelmy, Porter, Kilbourne, and
  Kelley}}]{beck07prl}
\bibinfo{author}{\bibfnamefont{B.~R.} \bibnamefont{Beck}},
  \bibinfo{author}{\bibfnamefont{J.~A.} \bibnamefont{Becker}},
  \bibinfo{author}{\bibfnamefont{P.}~\bibnamefont{Beiersdorfer}},
  \bibinfo{author}{\bibfnamefont{G.~V.} \bibnamefont{Brown}},
  \bibinfo{author}{\bibfnamefont{K.~J.} \bibnamefont{Moody}},
  \bibinfo{author}{\bibfnamefont{J.~B.} \bibnamefont{Wilhelmy}},
  \bibinfo{author}{\bibfnamefont{F.~S.} \bibnamefont{Porter}},
  \bibinfo{author}{\bibfnamefont{C.~A.} \bibnamefont{Kilbourne}},
  \bibnamefont{and} \bibinfo{author}{\bibfnamefont{R.~L.}
  \bibnamefont{Kelley}}, \bibinfo{journal}{\prl} \textbf{\bibinfo{volume}{98}},
  \bibinfo{pages}{142501} (\bibinfo{year}{2007}).

\bibitem[{\citenamefont{Tkalya et~al.}(2000)\citenamefont{Tkalya, Zherikhin,
  and Zhudov}}]{tkalya00prc}
\bibinfo{author}{\bibfnamefont{E.~V.} \bibnamefont{Tkalya}},
  \bibinfo{author}{\bibfnamefont{A.~N.} \bibnamefont{Zherikhin}},
  \bibnamefont{and} \bibinfo{author}{\bibfnamefont{V.~I.}
  \bibnamefont{Zhudov}}, \bibinfo{journal}{\prc} \textbf{\bibinfo{volume}{61}},
  \bibinfo{pages}{064308} (\bibinfo{year}{2000}).

\bibitem[{\citenamefont{Peik and \protect{Chr.} Tamm}(2003)}]{peik03epl}
\bibinfo{author}{\bibfnamefont{E.}~\bibnamefont{Peik}} \bibnamefont{and}
  \bibinfo{author}{\bibnamefont{\protect{Chr.} Tamm}}, \bibinfo{journal}{\epl}
  \textbf{\bibinfo{volume}{61}}, \bibinfo{pages}{181} (\bibinfo{year}{2003}).

\bibitem[{\citenamefont{Flambaum}(2006)}]{flambaum06prl}
\bibinfo{author}{\bibfnamefont{V.~V.} \bibnamefont{Flambaum}},
  \bibinfo{journal}{\prl} \textbf{\bibinfo{volume}{97}},
  \bibinfo{pages}{092502} (\bibinfo{year}{2006}).

\bibitem[{\citenamefont{Rosenband et~al.}(2008)\citenamefont{Rosenband, Hume,
  Schmidt, Chou, Brusch, Lorini, Oskay, Drullinger, Fortier, Stalnaker
  et~al.}}]{rosenband08sci}
\bibinfo{author}{\bibfnamefont{T.}~\bibnamefont{Rosenband}},
  \bibinfo{author}{\bibfnamefont{D.~B.} \bibnamefont{Hume}},
  \bibinfo{author}{\bibfnamefont{P.~O.} \bibnamefont{Schmidt}},
  \bibinfo{author}{\bibfnamefont{C.~W.} \bibnamefont{Chou}},
  \bibinfo{author}{\bibfnamefont{A.}~\bibnamefont{Brusch}},
  \bibinfo{author}{\bibfnamefont{L.}~\bibnamefont{Lorini}},
  \bibinfo{author}{\bibfnamefont{W.~H.} \bibnamefont{Oskay}},
  \bibinfo{author}{\bibfnamefont{R.~E.} \bibnamefont{Drullinger}},
  \bibinfo{author}{\bibfnamefont{T.~M.} \bibnamefont{Fortier}},
  \bibinfo{author}{\bibfnamefont{J.~E.} \bibnamefont{Stalnaker}},
  \bibnamefont{et~al.}, \bibinfo{journal}{\sci} \textbf{\bibinfo{volume}{319}},
  \bibinfo{pages}{1808} (\bibinfo{year}{2008}).

\bibitem[{\citenamefont{Hayes and Friar}(2007)}]{hayes07plb}
\bibinfo{author}{\bibfnamefont{A.~C.} \bibnamefont{Hayes}} \bibnamefont{and}
  \bibinfo{author}{\bibfnamefont{J.~L.} \bibnamefont{Friar}},
  \bibinfo{journal}{\plb} \textbf{\bibinfo{volume}{650}}, \bibinfo{pages}{229}
  (\bibinfo{year}{2007}).

\bibitem[{\citenamefont{Hayes et~al.}(2008)\citenamefont{Hayes, Friar, and
  M\"oller}}]{hayes08prc}
\bibinfo{author}{\bibfnamefont{A.~C.} \bibnamefont{Hayes}},
  \bibinfo{author}{\bibfnamefont{J.~L.} \bibnamefont{Friar}}, \bibnamefont{and}
  \bibinfo{author}{\bibfnamefont{P.}~\bibnamefont{M\"oller}},
  \bibinfo{journal}{\prc} \textbf{\bibinfo{volume}{78}},
  \bibinfo{pages}{024311} (\bibinfo{year}{2008}).

\bibitem[{\citenamefont{\protect{X.-t.} He and \protect{Z.-z.}
  Ren}(2008)}]{he08jpg}
\bibinfo{author}{\bibnamefont{\protect{X.-t.} He}} \bibnamefont{and}
  \bibinfo{author}{\bibnamefont{\protect{Z.-z.} Ren}}, \bibinfo{journal}{\jpg}
  \textbf{\bibinfo{volume}{35}}, \bibinfo{pages}{035106}
  (\bibinfo{year}{2008}).

\bibitem[{\citenamefont{Litvinova et~al.}(2009)\citenamefont{Litvinova,
  Feldmeier, Dobaczewski, and Flambaum}}]{litvinova09arxiv}
\bibinfo{author}{\bibfnamefont{E.}~\bibnamefont{Litvinova}},
  \bibinfo{author}{\bibfnamefont{H.}~\bibnamefont{Feldmeier}},
  \bibinfo{author}{\bibfnamefont{J.}~\bibnamefont{Dobaczewski}},
  \bibnamefont{and} \bibinfo{author}{\bibfnamefont{V.~V.}
  \bibnamefont{Flambaum}} (\bibinfo{year}{2009}),
  \bibinfo{note}{arXiv:0901.1240}.

\bibitem[{\citenamefont{Flambaum et~al.}(2009)\citenamefont{Flambaum, Auerbach,
  and Dmitriev}}]{flambaum09epl}
\bibinfo{author}{\bibfnamefont{V.~V.} \bibnamefont{Flambaum}},
  \bibinfo{author}{\bibfnamefont{N.}~\bibnamefont{Auerbach}}, \bibnamefont{and}
  \bibinfo{author}{\bibfnamefont{V.~F.} \bibnamefont{Dmitriev}},
  \bibinfo{journal}{\epl} \textbf{\bibinfo{volume}{85}}, \bibinfo{pages}{50005}
  (\bibinfo{year}{2009}).

\bibitem[{\citenamefont{Flambaum and Wiringa}(2009)}]{flambaum09prc}
\bibinfo{author}{\bibfnamefont{V.~V.} \bibnamefont{Flambaum}} \bibnamefont{and}
  \bibinfo{author}{\bibfnamefont{R.~B.} \bibnamefont{Wiringa}},
  \bibinfo{journal}{\prc} \textbf{\bibinfo{volume}{79}},
  \bibinfo{pages}{034302} (\bibinfo{year}{2009}).

\bibitem[{\citenamefont{Lide}(2009)}]{crchandbook}
\bibinfo{editor}{\bibfnamefont{D.~R.} \bibnamefont{Lide}}, ed.,
  \emph{\bibinfo{title}{CRC Handbook of Chemistry and Physics (Internet Version
  2009)}} (\bibinfo{publisher}{CRC Press/Taylor and Francis},
  \bibinfo{address}{Boca Raton, FL}, \bibinfo{year}{2009}),
  \bibinfo{edition}{89th} ed.

\bibitem[{\citenamefont{\protect{Segr\`e}}(1977)}]{segre77book}
\bibinfo{author}{\bibfnamefont{E.}~\bibnamefont{\protect{Segr\`e}}},
  \emph{\bibinfo{title}{Nuclei and Particles}}
  (\bibinfo{publisher}{Benjamin-Cummings}, \bibinfo{address}{Reading, MA},
  \bibinfo{year}{1977}).

\bibitem[{\citenamefont{Hasse and Myers}(1988)}]{hasse88book}
\bibinfo{author}{\bibfnamefont{R.~W.} \bibnamefont{Hasse}} \bibnamefont{and}
  \bibinfo{author}{\bibfnamefont{W.~D.} \bibnamefont{Myers}},
  \emph{\bibinfo{title}{Geometrical Relationships of Macroscopic Nuclear
  Physics}} (\bibinfo{publisher}{Springer-Verlag},
  \bibinfo{address}{Heidelberg}, \bibinfo{year}{1988}).

\bibitem[{\citenamefont{Angeli}(2004)}]{angeli04adndt}
\bibinfo{author}{\bibfnamefont{I.}~\bibnamefont{Angeli}},
  \bibinfo{journal}{\adndt} \textbf{\bibinfo{volume}{87}}, \bibinfo{pages}{185}
  (\bibinfo{year}{2004}).

\bibitem[{\citenamefont{Stone}(2005)}]{stone05adndt}
\bibinfo{author}{\bibfnamefont{N.~J.} \bibnamefont{Stone}},
  \bibinfo{journal}{\adndt} \textbf{\bibinfo{volume}{90}}, \bibinfo{pages}{75}
  (\bibinfo{year}{2005}).

\bibitem[{\citenamefont{Berengut et~al.}(2003)\citenamefont{Berengut, Dzuba,
  and Flambaum}}]{berengut03pra}
\bibinfo{author}{\bibfnamefont{J.~C.} \bibnamefont{Berengut}},
  \bibinfo{author}{\bibfnamefont{V.~A.} \bibnamefont{Dzuba}}, \bibnamefont{and}
  \bibinfo{author}{\bibfnamefont{V.~V.} \bibnamefont{Flambaum}},
  \bibinfo{journal}{\pra} \textbf{\bibinfo{volume}{68}},
  \bibinfo{pages}{022502} (\bibinfo{year}{2003}).

\bibitem[{\citenamefont{Dzuba et~al.}(1987)\citenamefont{Dzuba, Flambaum,
  Silvestrov, and Sushkov}}]{dzuba87jpb}
\bibinfo{author}{\bibfnamefont{V.~A.} \bibnamefont{Dzuba}},
  \bibinfo{author}{\bibfnamefont{V.~V.} \bibnamefont{Flambaum}},
  \bibinfo{author}{\bibfnamefont{P.~G.} \bibnamefont{Silvestrov}},
  \bibnamefont{and} \bibinfo{author}{\bibfnamefont{O.}~\bibnamefont{Sushkov}},
  \bibinfo{journal}{\jpb} \textbf{\bibinfo{volume}{20}}, \bibinfo{pages}{1399}
  (\bibinfo{year}{1987}).

\bibitem[{\citenamefont{Dzuba et~al.}(1996)\citenamefont{Dzuba, Flambaum, and
  Kozlov}}]{dzuba96pra}
\bibinfo{author}{\bibfnamefont{V.~A.} \bibnamefont{Dzuba}},
  \bibinfo{author}{\bibfnamefont{V.~V.} \bibnamefont{Flambaum}},
  \bibnamefont{and} \bibinfo{author}{\bibfnamefont{M.~G.}
  \bibnamefont{Kozlov}}, \bibinfo{journal}{\pra} \textbf{\bibinfo{volume}{54}},
  \bibinfo{pages}{3948} (\bibinfo{year}{1996}).

\bibitem[{\citenamefont{Dzuba and Johnson}(1998)}]{dzuba98pra}
\bibinfo{author}{\bibfnamefont{V.~A.} \bibnamefont{Dzuba}} \bibnamefont{and}
  \bibinfo{author}{\bibfnamefont{W.~R.} \bibnamefont{Johnson}},
  \bibinfo{journal}{\pra} \textbf{\bibinfo{volume}{57}}, \bibinfo{pages}{2459}
  (\bibinfo{year}{1998}).

\bibitem[{\citenamefont{Blaise and Wyart}(1992)}]{actinidesbook}
\bibinfo{author}{\bibfnamefont{J.}~\bibnamefont{Blaise}} \bibnamefont{and}
  \bibinfo{author}{\bibfnamefont{J.-F.} \bibnamefont{Wyart}},
  \emph{\bibinfo{title}{Selected Constants: Energy Levels and Atomic Spectra of
  Actinides}} (\bibinfo{publisher}{Tables Internationales de Constantes},
  \bibinfo{address}{Paris}, \bibinfo{year}{1992}).

\end{thebibliography}

\end{document}